\documentclass[twocolumn,superscriptaddress,secnumarabic,showpacs, amssymb, nobibnotes, aps, bm,prb,floatfix,longbibliography]{revtex4-2}

\usepackage{amsmath}
\usepackage{amssymb}
\usepackage{graphicx}
\usepackage{verbatim}
\usepackage{braket}
\usepackage{ulem}
\usepackage{soul}
\usepackage{gensymb}
\usepackage[usenames]{color}

\renewcommand\vec{\boldsymbol}

\begin{document}

\title{
 Competition of Exchange and Correlation Energies\\ 
in Two-Dimensional $N$-component Electron Gas Ferromagnetism }

\author{Chen-How Huang}

\affiliation{Donostia International Physics Center (DIPC), 20018 Donostia--San Sebasti\'an, Spain}
\affiliation{Departamento de Polímeros y Materiales Avanzados: Física, Química y Tecnología, Facultad de Ciencias Químicas,
Universidad del País Vasco UPV/EHU, 20018 Donostia-San Sebastián, Spain.}

\author{ Chunli Huang}
\affiliation{Department of Physics and Astronomy, University of Kentucky, Lexington, Kentucky 40506-0055, USA}

\author{M. A. Cazalilla}
\affiliation{Donostia International Physics Center (DIPC), 20018 Donostia--San Sebasti\'an, Spain}
\affiliation{IKERBASQUE, Basque Foundation for Science, Plaza Euskadi 5
48009 Bilbao, Spain}

\date{\today} 

\begin{abstract}
Motivated by recent observations of symmtry broken phases in lightly-doped multilayer graphene, we investigate magnetic phase transitions in a generalized electron gas model with four-component electron spin. This model simplifies the problem with a parabolic dispersion band, abstracting away the details of the graphene band structure to focus solely on the effects of the Coulomb interaction. We report four findings: 1) In the Hartree-Fock approximation, we observe that the paramagnetic state undergoes a sequence of density-driven, first-order phase transitions, progressively depopulating electrons from each spin component until achieving complete polarization within a very narrow density window where $1.2<r_s<2$ ($r_s$ being the electron gas parameter). 2) Further incorporating the correlation energy via the Bohm-Pines random-phase approximation shows that the cascade of transitions obtained within Hartree-Fock approximation is replaced by a single ferromagnetic phase transition at $r_s = 6.12$.
3) The disappearance of cascade is due to the correlation energy difference between the four-component paramagnetic state and symmetry-broken phases, which is nearly an order of magnitude more negative than the corresponding Hartree-Fock energy difference for  $1.2 < r_s < 2$.
4) The transition from the paramagnetic state to the fully polarized state at $r_s=6.12$ is governed by the balance between exchange and correlation energies, a competition that cannot be captured by mean-field approximations to models featuring effective (density-dependent) delta-function interactions, such as the Stoner model. We use the insights from our model to comment on the phase diagram of multilayer graphene electron gas.

\end{abstract}

\maketitle

\section{Introduction}
The recent discovery of magnetic metals and superconductivity in lightly-doped Bernal bilayer graphene~\cite{zhou2022isospin} and Rhombohedral trilayer 
graphene~\cite{zhou2021half,zhou2021superconductivity,seiler2022quantum}
provides an exciting opportunity to study strongly-correlated physics in one of the most tunable 
two-dimensional materials. The preliminary understanding of the phase diagram is based on the Hartree-Fock mean-field theory computed using an experimentally-informed band structure and a modified Coulomb interaction that accounts for the gate screening effect \cite{chatterjee2022inter,huang2023spin,koh2024correlated,zhou2021half}. While mean-field theory elucidates the nature of the symmetry broken magnetic phases, it fails to explain the re-entrance of the symmetric paramagnetic phase around the half-metal in Bernal bilayer graphene. In trilayer graphene, it also wrongly predicts that the magnetic transition persists all the way down to zero displacement field whenever the density is close to the Van Hove singularity. These disagreements call for a detailed study on the importance of correlation energy in the phase diagram. The band structure of bilayer and trilayer graphene contains several important details that are all very important. 
Firstly, the density-of-states of the normal state is four-fold degenerate because of the spin and valley degrees of freedom. Secondly, the density of states of the first valence and conduction bands is strongly enhanced by the electric displacement field and proximity to  Van Hove singularities. Thirdly, the topology of the Fermi sea at these Van Hove singularities changes from a simply-connected Fermi sea to an annular shape and from an annular shape to three disjoint Fermi pockets. In order to have a thorough understanding of the experimental phase diagram, it is important to consider simpler models that capture the above band structure details one at a time.

In this article, we study ferromagnetic phase transitions of a \textit{jellium} model 
for a two-dimensional Fermi gas with $N=4$ (spin) components. The generalization from the standard $N=2$ to $N=4$ is intended to mimic the four-fold spin-valley degeneracy of graphene. This approach is informed by previous studies in multi-valley semiconductors \cite{PhysRevB.68.205322_DFT, rasolt1990continuous}.
In the Hartree-Fock  approximation (HFA), we identify a sequence of first-order magnetic phase transitions. Starting from the paramagnetic state as the ground state at large densities, it sequentially transitions into the so-called three-quarter metal at $r_s=1.2$,  half-metal at $r_s=1.5$, and  quarter metal at $r_s=2$. Of all these phases, the three-quarter metal state emerges as the ground state for the smallest density interval, being tightly flanked by the more stable paramagnetic and half-metal states. This narrow stability range offers a potential explanation for the absence of the three-quarter metal phase in multilayer graphene, as found in experiments
\cite{zhou2021half,zhou2021superconductivity,zhou2022isospin}. On the other hand, 
in the Bohm-Pines random phase approximation (RPA),  the partially polarized three-quarter or  half-metal never become the ground state. This is because the RPA correlation energy of the paramagnetic state is significantly lower than the partially polarized state in the density range where $1.2<r_s<2$ for which the HFA yields a cascade of magnetic transitions.
Improvements beyond the RPA are unlikely to reinstate the two partially polarized states as the ground state due to the large RPA correlation energy difference between the paramagnetic state and the partially polarized state for $1.2 <r_s<2$. This seems to suggest that the observed half-metal in multilayer graphene is likely  stabilized by combination of other important factors, such as density of states and/or the topology of the bands, which our simple model does not account for. 

This article is organized as follows.
In section \ref{sec:HF}, we present the analysis using the Hartree-Fock approximation. In section~\ref{sec:RPA}, we expand on this by incorporating correlation energy through the random phase approximation (RPA) to examine its impact on the transitions. In section ~\ref{sec:stoner_model}, we compare the RPA results and the mean-field approximation in the Stoner model, where the conventional Coulomb potential is substituted with a delta function potential with a density dependent interaction parameter. In the conclusion section~\ref{sec:fin} our results are summarized and we comment on the implications of our work for multilayer graphene.

\section{Model and Hartree-Fock Approximation}\label{sec:HF}

The model explored in this study is an extension of the standard \textit{jellium} model to two spatial dimensions, accommodating $N$ spin-components, where $N$ is an arbitrary integer. The Hamiltonian, exhibiting SU($N$) symmetry, is given by the following:
\begin{align}
    H = \sum_{\vec{k}} \epsilon_{\vec{k}} c^{\dag}_{\vec{k}\sigma}c_{\vec{k}\sigma}  +
    \tfrac{1}{2\Omega}\sum_{\vec{q}\neq \vec{0}} V_{\vec{q} }\rho_{\vec{q}} \rho_{-\vec{q}} \label{eq:ham}
\end{align}
where $\epsilon_{\vec{k}} = \frac{\hbar^2\vec{k}^2}{2m}$ is the free-fermion dispersion,
$c^{\dag}_{\vec{k}\sigma}$  ($c_{\vec{k}\sigma}$) creates (destroys) an electron in a plane-wave state with wave vector $\vec{k}$ and SU($N$) quantum number $\sigma=1,\ldots, N$. The occupation of the
single-electron states is given by 
the operator $n_{\vec{k}}=\sum_{\sigma}
c^{\dagger}_{\vec{k},\sigma}c_{\vec{k},\sigma}$. 
The operator $\rho_{\vec{q}} = \sum_{\vec{k},\sigma} c^{\dagger}_{\vec{k+q},\sigma}c_{\vec{k},\sigma}$ is the Fourier components of the electron density with wave vector $\vec{q}$. 
The electron-electron interaction (in Fourier components) is described by the \textit{bare} Coulomb potential, i.e.
$V_{\vec{q}}=2\pi e^2/|\vec{q}|$; $\Omega$ is the area of the system.

In the Hartree-Fock approximation (HFA), we assume the ground state is a Slater-determinant, parameterized by a set of Fermi radii $\{ k_{F\sigma} \}$ for $\sigma = 1$ to $N$. In second quantization, this wavefunction  reads

\begin{equation}
    |\text{HF} \rangle =  \prod_{\sigma=1}^{N} \prod_{|\vec{k}|<k_{F\sigma}} c_{\vec{k},\sigma}^{\dagger} |0\rangle 
\end{equation}

Upon calculating the expectation value of the Hamiltonian $H$ got this state,  the energy per particle is  function of the filling fractions $\nu_\sigma$ takes the form:
\begin{align} \label{eq:tot_enrg_HF}
    \epsilon_{HF} &= \frac{\langle \text{HF}|H | \text{HF}\rangle}{N_{e} } \nonumber \\
    &= \frac{2}{r_s^2}\sum_{\sigma} \nu_\sigma^2 - \frac{16}{3\pi r_s} \sum_{\sigma}\nu_{\sigma}^{3/2},
\end{align}
where the filling fraction $\nu_\sigma$ is defined as:
\begin{equation}
    \nu_\sigma =\frac{k_{F\sigma}^2}{\sum_{\sigma} k^2_{F\sigma}}.
\end{equation}
Here, $N_{e}=\sum_{\vec{k},\sigma} \theta(k_{F\sigma}-|\vec{k}|)$ represents the total number of electrons. The parameter $r_s$, known as the electron gas parameter, is defined by the relation $ \pi(r_s a_B)^2=1/n_e$, where $a_B$ is the Bohr radius and $n_e$ is the total electron density.

Next, we minimize Eq.~\eqref{eq:tot_enrg_HF} with respect to $\nu_\sigma$ at different $r_s$ subject to the condition $\sum_{\sigma=1
}^N\nu_{\sigma}=1$, which fixes the total density. This optimization resulted in multiple distinct solutions, which we classify using an $N$-component vector $\vec{\nu}_j^N = (\nu_1,\nu_2,\cdots,\nu_j,0,\cdots,0)$. For instance, when $\nu_1=\nu_2=...=\nu_j=1/j$,  there are $j$ equally-populated components,  while the remaining $N-j$ components are empty. Consequently, this configuration leads to a reduction in the $SU(N)$ symmetry of the Hamiltonian (Eq.~\eqref{eq:ham}) to $SU(j)\times SU(N-j)$, effectively serving as an order parameter by reflecting the ordering that results from the biased population of the components.

For the specific case of $N=4$, which is pertinent to multilayer graphene's spin and valley degree of freedom, we have assigned the following nomenclature: $\vec{\nu}_{j=1}^{N=4}$ as the fully-polarized metal, $\vec{\nu}_{j=2}^{N=4}$ as the half-metal, $\vec{\nu}_{j=3}^{N=4}$ as the three-quarter metal and $\vec{\nu}_{j=4}^{N=4}$ as the fully symmetric metal (or paramagnetic state). These classifications are further elaborated in table \ref{tab:SU(4)}.

\begin{table}[t!]
  \begin{center}
    \begin{tabular}{l|c|c}
      \textbf{notation} & \textbf{filling fractions}&
      \textbf{states}\\
      \hline
      $\vec{\nu}^{N=4}_{4}$ & $(1/4,1/4,1/4,1/4)$ &\text{Paramagnet} \\
      $\vec{\nu}^{N=4}_{3}$ & $(1/3,1/3,1/3,0) $&\text{Three quarter metal} \\
      $\vec{\nu}^{N=4}_{2}$ & $(1/2,1/2,0,0) $ &\text{Half-metal}\\
      $\vec{\nu}^{N=4}_{1}$ & $(1,0,0,0) $ &\text{Fully-polarized metal}\\
    \end{tabular}
  \end{center}
  \caption{
  Definitions of $\nu^{N=4}_j$ for the four competing ground states in an $SU(4)$ system.}\label{tab:SU(4)}
\end{table}

\begin{figure}[t]
\includegraphics[width=\columnwidth]{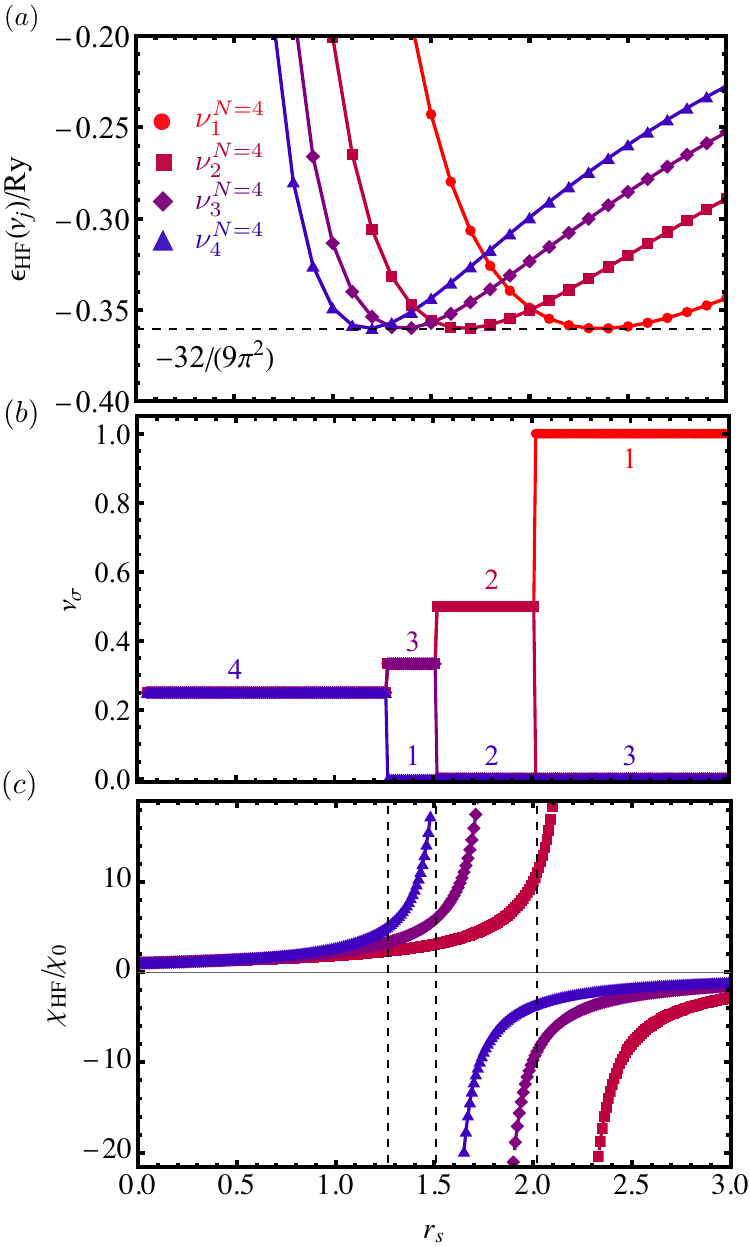}
\center
\caption{(a) $\epsilon_{HF}$ versus $r_s$ in the two-dimensional electron gas with SU$(4)$ symmetry.  All energy curves share the same energy minimum of $-32/9\pi^2$ shifting towards larger $r_s$ as the number of occupied Fermi surfaces decreases. (b) Ground-state order parameter $\vec{\nu}_{\sigma}^{N=4}$ versus~ $r_s$. The numbers indicate the multiplicity of overlapping lines. (c) Spin susceptibility $\chi_{HF}$ for filling fractions $\vec{\nu^4_2},\vec{\nu^4_3}$ and $\vec{\nu^4_4}$ states (see table~\ref{tab:SU(4)}) calculated using Hartree-Fock approximation. The dashed vertical line marks the point of energy level crossing, highlighting the significantly enhanced spin susceptibility that occurs  prior to this transition. }\label{fig:HF}
\end{figure}

Fig.~\ref{fig:HF}(a) shows the curves of the total HF energy versus $r_s$ of the four competing magnetic states for the $N=4$ electron gas. The transitions between these states occur at the points where two energy curves intersect, indicating a discontinuity in chemical potential and a change in the order parameter $\vec{\nu}^{N=4}_{j}$, shown in  Fig.~\ref{fig:HF}(b). As $r_s$ increases, the system progresses from the paramagnetic state $\vec{\nu}^4_4$ through a series of first-order phase transitions: $\vec{\nu}^4_4 \to \vec{\nu}^4_3 \to \vec{\nu}^4_2 \to \vec{\nu}^4_1 $, sequentially depopulating one component at a time until reaching full polarization. 
An important feature of the energy curves in Fig.~\ref{fig:HF}(a) is that they all share the same absolute energy minimum and reduction of the number of Fermi surface merely shifts the curve towards to larger $r_s$. To understand this trend, we write Hartree-Fock (HF) energy as $\epsilon_{HF}=a(\vec{\nu})/r_s^2-b(\vec{\nu})/r_s$. The stationary equation, $d\epsilon_{HF}/dr_s=0$, yields the solution:
\begin{equation}
    r_s^* = \frac{2a(\vec{\nu})}{b(\vec{\nu})}\;,\; \epsilon_{HF}(r_s^*)= -\frac{b^2(\vec{\nu})}{4a(\vec{\nu})}.
\end{equation}
Here $b(\vec{\nu})=\frac{16}{3\pi}\sum_{\sigma}\nu_{\sigma}^{3/2}$ and $a(\vec{\nu})=2\sum_{\sigma}\nu_\sigma^2$. When the filling fraction corresponds to a configuration with $j$ equally populated components, we find $b=\frac{16}{3\pi\sqrt{j}}$ and $a=2/j$.  Consequently, this results in $\epsilon_{HF}(r_s^*)= -32/9\pi^2$, a minimum energy value that remarkably does not depend on $j$. This result explains why, in the occupied manifold, each component has an equal electron count of $1/j$, rather than a skewed distribution where some components are more populated than others. We will subsequently use the energy increase associated with such skewed distributions to compute the susceptibility.

We would also like to point out that the density-range where $\vec{\nu}^4_3$ is the ground state is the smallest because the ground state energy of $\vec{\nu}^4_4 $ and $\vec{\nu}^4_2$ are very close by. This pattern of depopulating one component at a time agrees with mean-field calculations in multilayer graphene, where, as density approaches neutrality, the system transitions from a paramagnetic state to a three-quarter metal, then to a half-metal, and finally to a quarter metal. 
This symmetry breaking pattern contrasts sharply with the behavior observed under short-range delta-function interactions, where the system transitions directly from a fully symmetric state to a fully polarized state as density increases \cite{Huang_2023}.

Motivated by the experimental observation of superconductivity in multilayer graphene near magnetic-phase boundaries, which may be driven by ferromagnetic fluctuations \cite{zhou2021superconductivity,crepel2022spin, huang2022pseudospin, cea2023superconductivity}, it is valuable to compute the spin susceptibility. The generalized (uniform) spin susceptibility, which reflects changes in the order parameter within the occupied  $SU(j)$,  can be determined by computing the increase in HF energy in response to a small change in the filling fractions  ($\delta\ll 1$):
\begin{align}
\delta \epsilon_{HF}
=&\epsilon_\text{HF}\left[\nu_1+\frac{\delta}{j-1},\nu_2+\frac{\delta}{j-1},...,\nu_j-\delta,0,\cdots,0 \right],
\end{align}
From second derivative of the above energy variation with respect to $\delta$, we obtain the inverse paramagnetic susceptibility, which is independent of the sign of $\delta$:
\begin{align}
\frac{1}{\chi_{HF}}
&=\frac{1}{j-1}\sum_\sigma\frac{\partial^2 \epsilon_{\text{tot}}}
{\partial {\nu_\sigma}^2}\bigg|_{\vec{\nu}_j^N},\notag\\
&-\left(\frac{1}{j-1}\right)^2\sum_{\substack{\sigma\,\sigma'\\ \sigma\neq\sigma'}}\frac{\partial^2 \epsilon_\text{tot}}
{\partial {\nu_\sigma}\partial{\nu_{\sigma'}}}\bigg|_{\vec{\nu}_j^N}.
\end{align}
where $\epsilon_{\text{tot}} = \epsilon_{HF}$ in the HFA.
Fig.~\ref{fig:HF}(c) shows the susceptibilities normalized by $\chi_0= 4/r_s^2$, which is the susceptibility calculated for four-component non-interacting electron gas, (i.e.~density-of-states per particle).  The vertical lines indicate the positions where the energy-level crossings occur (first order phase transitions).
The figure shows that the susceptibilities diverge at an $r_s$ slightly larger than the energy level crossings, indicating that prior to depopulating a flavor at the first-order phase transition, there is an strongly enhanced spin susceptibility. This behavior is similar to the enhanced spin susceptibility observed in He$^3$ near its melting curve. Notably, such an enhanced spin susceptibility is crucial for stabilizing the Anderson-Morel anisotropic superfluid. Importantly, since long-wavelength spin fluctuation-mediated superconductivity can occur not only near second-order phase transitions but also adjacent to first-order transition boundaries, where susceptibility is markedly increased, the enhanced spin susceptibility can substantially influence the bare Coulomb repulsion in multilayer graphene.

\section{Correlation Energy}\label{sec:RPA}

It is well documented that the HFA inaccurately describes the energy differences between the ferromagnetic and the 
paramagnetic states of the electron gas~\cite{giuliani_vignale_2005, rado1966exchange} at intermediate $r_s$. In this section, we explore the impact of  the correlation energy estimated within the random-phase approximation (RPA)
on ferromagnetic transitions. 
It is worth recalling that  
the exact correlation energy per particle can
be written in terms of the equal time density-density correlation function using the method of coupling-constant  integration~\cite{giuliani_vignale_2005} as follows:
\begin{equation}
    \epsilon_c =    \frac{1}{2\Omega}\sum_{\vec{q}\neq\vec{0}} \frac{V_q}{N_e}  \int_0^1 d\lambda \langle \psi_\lambda |  \rho_{\vec{q}} \rho_{\vec{-q}}| \psi_\lambda \rangle  - \epsilon_{HF},
\end{equation}
where $\epsilon_{HF}$ is given by Eq.~\eqref{eq:tot_enrg_HF}.  The RPA correlation energy per particle measured in Ry for a state with filling factors $\vec{\nu}$ is given by the following expression~\cite{giuliani_vignale_2005}:
\begin{equation} \label{eq:epsilon_c}
    \epsilon_{c}(\vec{\nu}) = 
\pi(a_Br_s)^2\int \frac{d^2q}{(2\pi)^2}  W(q;\vec{\nu}),\\
\end{equation}
where
\begin{align}\label{eq:W}
     W(q;\vec{\nu}) &= \frac{2}{\beta}  \sum_{\omega_n}  \ln \left[1-V_q \Pi_{q,\omega_n}(\vec{\nu})\right] \notag 
     \\ &\qquad +  \frac{2}{\beta}  \sum_{\omega_n}  V_q\Pi_{q,\omega_n}(\vec{\nu})  .
\end{align}
Here $W(q;\vec{\nu})$ is the correlation energy at wavenumber $q=|\bf{q}|$ and  $\Pi_{q,\omega_n}$ is the particle-hole susceptibility  (Lindhard  function) of the ground state with filling fractions $\vec{\nu}$. Explicitly, the Lindhard function at zero temperature in two-dimensions is
\begin{align}\label{eq:pi}
    \Pi_{q,\omega_n}(\vec{\nu}) &= \sum_{\sigma} \int  \frac{d^2k }{(2\pi)^2} \frac{ 
    n_{F\sigma}(\epsilon_{\vec{k+q}};\vec{\nu})-n_{F\sigma}(\epsilon_{\vec{k}};\vec{\nu})}{i\omega_n+\epsilon_{\vec{k+q}}-\epsilon_{\vec{k}}}
    \notag\\
    &=\sum_{\sigma} \phi\left[z_-,q,k_{F,\sigma}(\vec{\nu})\right]-\phi\left[z_+,q,k_{F,\sigma}(\vec{\nu})\right].
\end{align}
where $n_{F\sigma}(\epsilon;\vec{\nu})=\Theta\left[ \mu_\sigma(\vec{\nu})-\epsilon\right]$ is the Fermi-Dirac distribution at zero temperature, $\Theta(x)$ being the Heaviside step-function; $\epsilon_{\vec{k}}=\hbar^2\vec{k}^2/2m$, $\mu_\sigma(\vec{\nu})=\hbar^2k^2_{F,\sigma}(\vec{\nu})/2m$  the free-electron dispersion and  $k_{F,\sigma}(\vec{\nu})$ is the Fermi momentum for the $\sigma$-component as determined from the 
filling-fraction vector $\vec{\nu}$; $\omega_n=2\pi n/\beta$ are the boson Matsubara frequencies and  $\phi$  is a function defined by the expression:
\begin{align}\label{eq:phi}
    \phi(z,q,k_F)&=\frac{2m\pi}{\hbar^2} \left( \frac{k_F}{2\pi} \right)^2\notag\\
    \times&
\frac{1}{k_F^2 q^2}\left[ z-\left( z+k_Fq \right)\sqrt{\frac{z-k_Fq}{z+k_Fq}}\: \right]
\end{align}
where $z_\pm=i m \omega/\hbar^2 \pm q^2/2$.

\begin{figure}[t]
\includegraphics[width=\columnwidth]{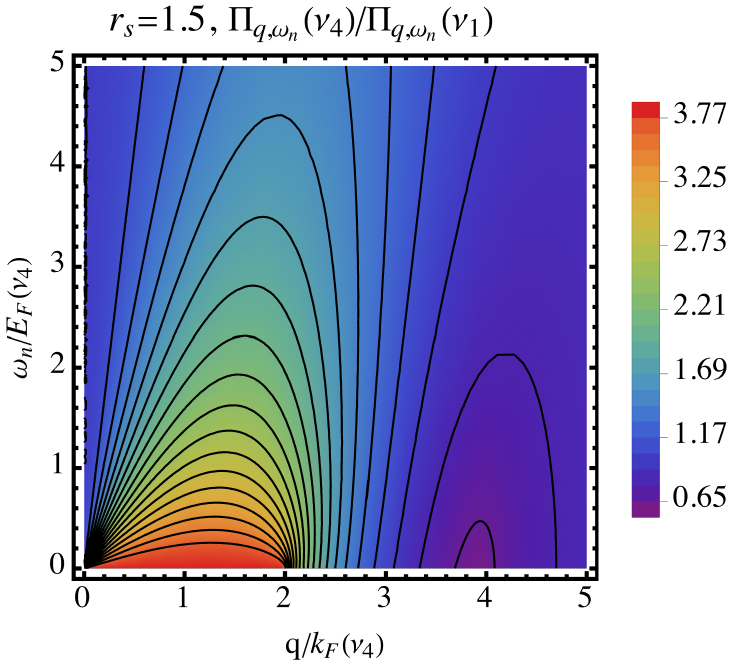}
\center
\caption{The ratio of (zero-temperature) particle-hole susceptibilities $\Pi_{q,\omega_n}(\vec{\nu}_4)/\Pi_{q,\omega_n}(\vec{\nu}_1)$ compares the paramagnetic state with four small Fermi surfaces to the quarter metal with one large Fermi surface in the $q$-$\omega_n$ plane. Notably, the  susceptibility of the paramagnetic state is significantly enhanced, particularly at low imaginary frequencies and small  wavevectors (up to approximately $2k_F$);  $k_F= k_F(\vec{\nu}_4) =1/(r_sa_B)$ is the Fermi wavenumber for the (one-component) quarter-metal and $E_F= \hbar^2k_F^2/2m$
}\label{fig:chi}
\end{figure}

At constant electron density (or $r_s$), increasing the number of degenerate Fermi surfaces leads to a decrease in the chemical potential, thereby enhancing the particle-hole susceptibiity. Consequently, a paramagnetic state with four Fermi surfaces exhibits a larger particle-hole susceptibility than symmetry-broken states with fewer Fermi surfaces. This effect is shown in Fig.~\ref{fig:chi}, where we observe an increase in the particle-hole susceptibility in most of the  $q$-$\omega_n$ plane except in regions where $q\geq 4k_F(\vec{\nu}_4)$. For an electron gas with parabolic dispersion, the susceptibility in the $j$-component paramagnetic state is related to the single-component ferromagnetic state as follows: $\Pi_{q,\omega_n}(\vec{\nu}_j)=j\Pi_{q\sqrt{j},j\omega_n}(\vec{\nu}_{j=1})$.

\begin{figure}[t]
\includegraphics[width=\columnwidth]{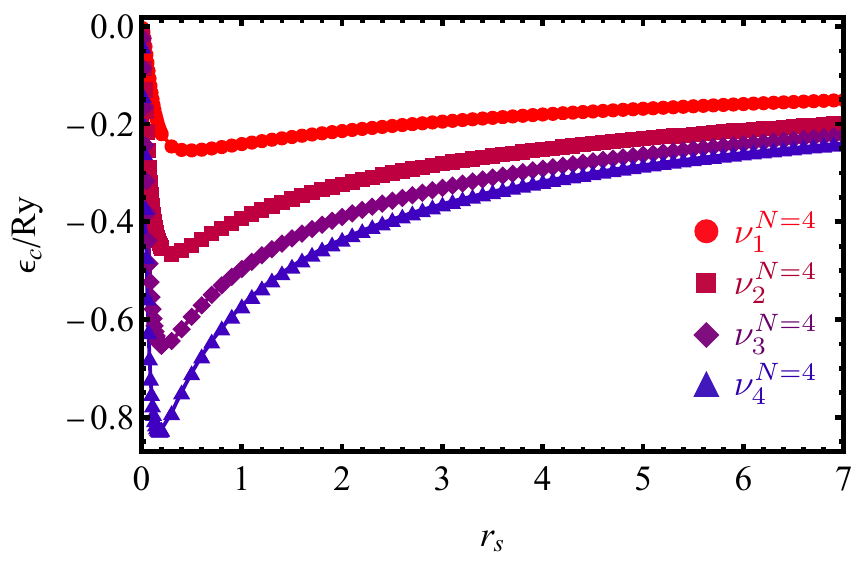}
\center
\caption{Correlation energy versus $r_s$ for the four competing states.  The correlation energy of the paramagnetic state $\vec{\nu}_{4}^{N=4}$ is lower than for the three-quarter metal $\vec{\nu}_{3}^{N=4}$, which is in turn lower than for the half-metal $\vec{\nu}_{2}^{N=4}$,
and the quarter metal $\vec{\nu}_{1}^{N=4}$ has the highest correlation energy among the states shown.
}\label{fig:correlation_energy}
\end{figure}

\begin{figure}
 \includegraphics[width=\columnwidth]{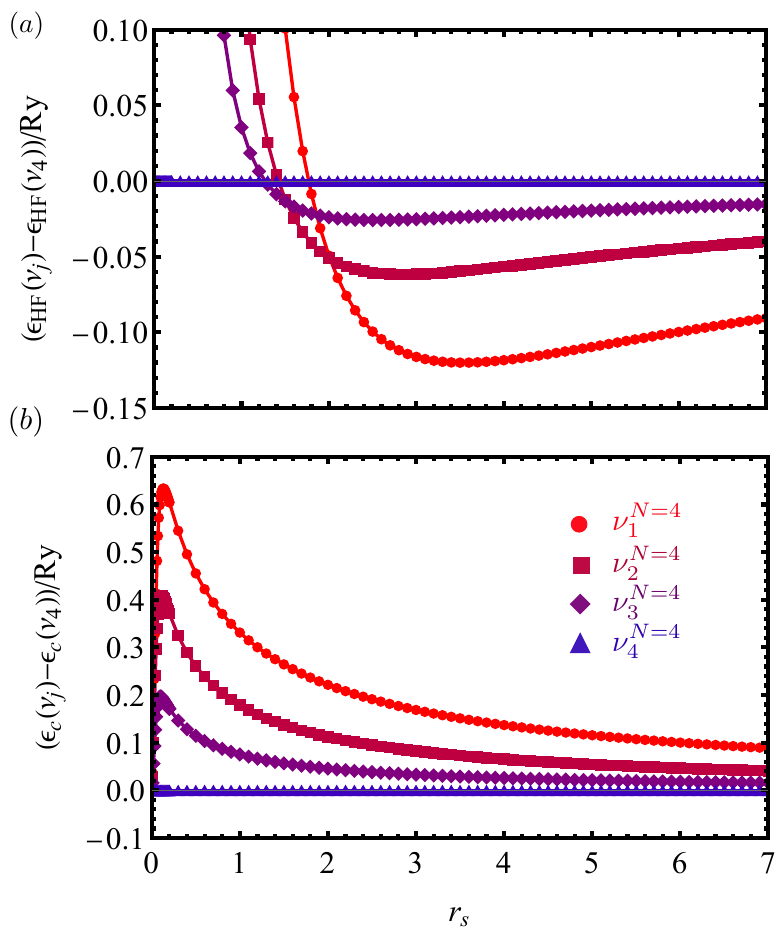}
\center
\caption{(a) Hartree-Fock energy differences between the three symmetry-broken states and the paramagnetic state  $\vec{\nu_4}$.  (b) Correlation energies between the three symmetry-broken states and the paramagnetic state  $\vec{\nu}^{N=4}_4$. Note the difference in the y-axis scale especially in the regin of $1<r_s<2$.
 }\label{fig:energy_diff}
\end{figure}

Due to the increase in magnitude of the particle-hole susceptibility in the $q$-$\omega_n$ plane with the number of degenerate Fermi surfaces $j$, the correlation energy, which is  sum the geometric series of $V_q \Pi_{q,\omega_n}(\vec{\nu})$,  becomes more negative as $j$ increases.
This trend is illustrated in Fig.~\ref{fig:correlation_energy}. 
It is important to emphasize that while the HF energy, as shown in Fig.~\ref{fig:HF}(a), and the correlation energy, as shown in Fig.~\ref{fig:correlation_energy}, are of the same order of magnitude, the HF energy differences between the symmetry-broken  and the paramagnetic states are an order of magnitude smaller than the corresponding differences in correlation energy for $1<r_s<2$. This significant disparity is also shown in Fig.~\ref{fig:energy_diff}.

\begin{figure}
\includegraphics[width=\columnwidth]{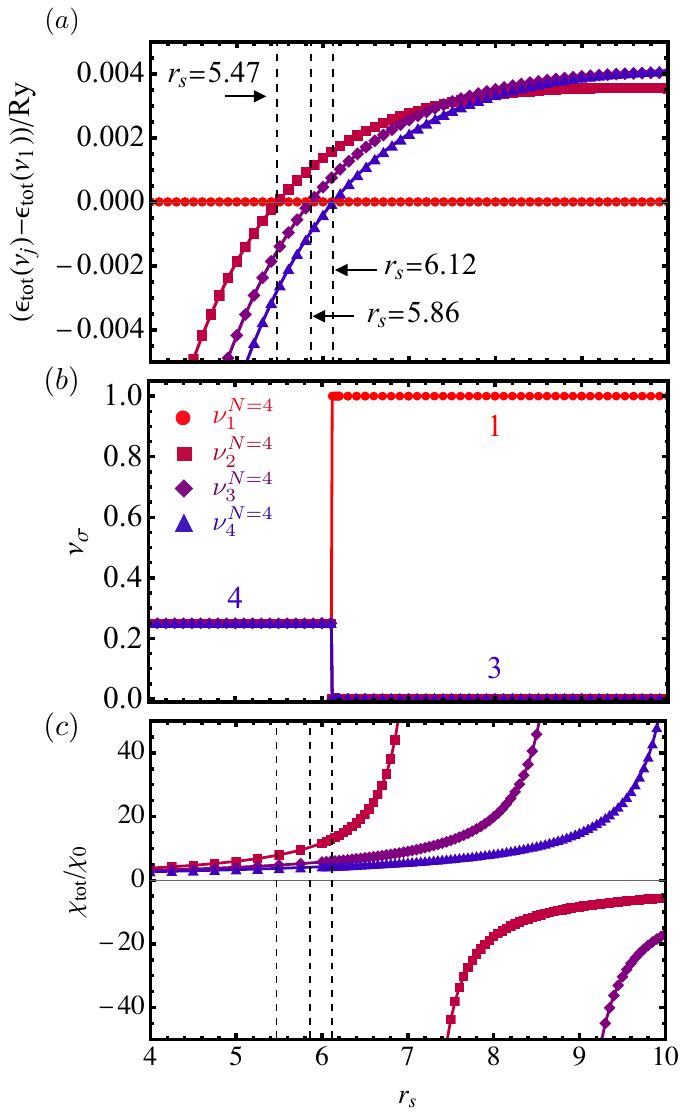}
\center
\caption{ (a) Total energies per particle relative to the fully polarized state calculated using the random phase approximation (RPA) for the four states listed in table~\ref{tab:SU(4)}. The dash lines indicate the transition points from the paramagnetic states with $N \leq 4) $ (i.e. $\vec{\nu}^{N=2}_2,\vec{\nu}^{N=3}_3$ and $\vec{\nu}^{N=4}_4$) to the fully polarized state, $\vec{\nu}^{N=2,3,4}_1$  according to the energetic  relation Eq.\eqref{eq:theorem}. (b) Behavior of tilling fraction of each SU($4$)-spin component in two dimensions versus $r_s$ showing the ferromagnetic transition. The numbers indicate the multiplicity of overlapping lines. (c) Susceptibilities calculated from total energy including the correlation energy calculated using the RPA. The dashed lines correspond to the ferromagnetic transition for  $N=2,3$ and $4$ component systems. Divergences are observed at higher values than those where the transition from paramagnet to (fully polarized) ferromagnet  occurs. Compared to the results from the Hartree-Fock approximation,  a suppression on the susceptibilities with increasing number of component is observed. 
 }\label{fig:RPA}
\end{figure}

Fig.~\ref{fig:RPA}(a)  shows the  total energies of the magnetic states relative to the fully polarized state as a function of $r_s$ where $ \epsilon_{tot}= \epsilon_c+ \epsilon_{HF}$. We found that including the RPA correlation energy alters the picture of the cascade of  phase transitions obtained from the HFA. Thus, a single first-order phase transition from the paramagnetic state (corresponding to the blue line in Fig.~\ref{fig:RPA}(a)) to the fully-polarized state (shown as the red-horizontal line) at $r_s=6.12$ occurs. 
The observed behavior, which sees the partially polarized states (i.e.  the half and three-quarter metal) pushed  up in energy  above fully polarized state, is also observed for other values of $N$. Indeed, using the result from  Eq.~\eqref{eq:theorem},
we have located the values of $r_s$ of the first-order phase transition from paramagnetic to fully-polarized state for the electron gases with $N=2$ and $N=3$ components, which happen for  $r_s=5.47$ and  $r_s=5.86$, respectively, and are indicated by the vertical dashed lines on Fig.~\ref{fig:RPA}(a)
Thus, within the RPA, $N$-component electron gas  exhibits a single phase transition from the  paramagnetic state to a  fully-polarized state. The  critical $r_s$ where the phase transition happens increases with the number of components $N$.
In  Fig.~\ref{fig:RPA}(c) we have plotted the susceptibility computed   as described in the previous section for different ground states characterized by different filling fractions $\vec{\nu}$. Including RPA corrections, we find a suppression relative to the HFA as the number of spin components $N$ becomes larger. Again, the dashed lines indicate the transition points for $N=2$, $3$ and $4$ component systems. Similarly  to the HF susceptibility, divergences for the partially polarized states are observed for values of  $r_s$ larger than the value of the first-order ferromagnetic transition (in this case to the fully polarized state).

\begin{figure}[t]
\includegraphics[width=\columnwidth]{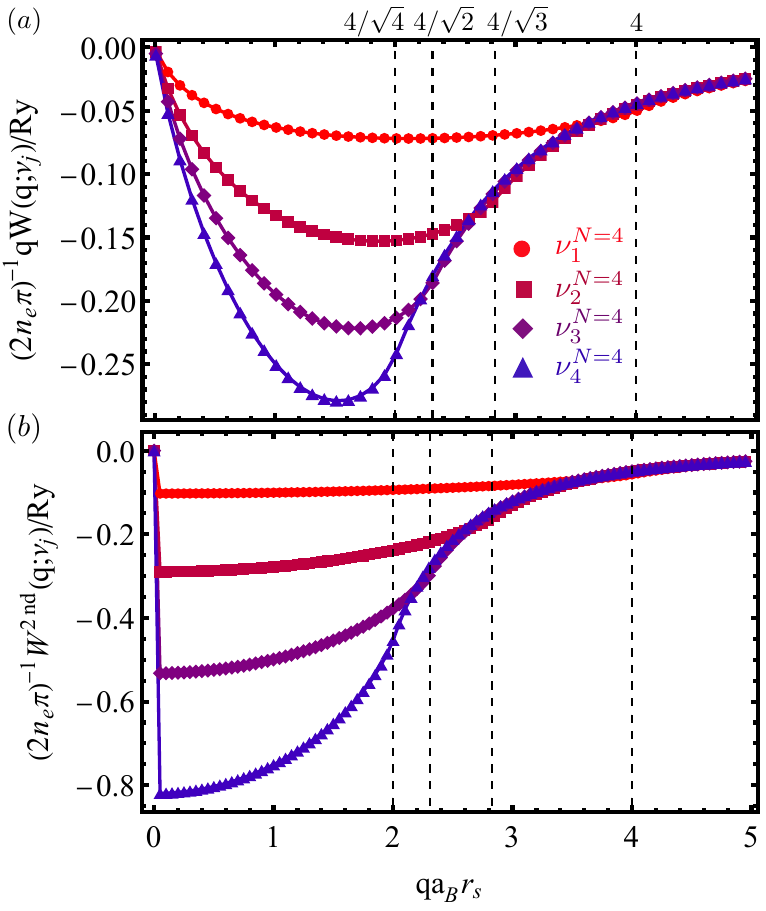}
\center
\caption{  
(a) $qW(q;\vec{\nu}^{N=4}_j)/(2n_e\pi)$ calculated from the RPA for the four states in table~\ref{tab:SU(4)} at $r_s=1.5$.  The area under each curve quantifies the correlation energies of each state, with a noted increase in magnitude for systems having a larger number of components. (b) The function $qW(q;\vec{\nu}^4_j)/(2n_e\pi)$, calculated using second-order perturbation theory for the same states at $r_s=1.5$, closely matches the RPA results at higher $q$ values. The $x$-axis is labeled as $qa_Br_s$ and for each state, the $2k_F(\vec{\nu}^N_j)$ position is indicated by a vertical dashed line located at $4/\sqrt{j}$ where $j$ ranges from 1 to 4.
}\label{fig:correlation_energy_W} 
\end{figure}

 Further understanding can be gained by studying the function $q W(q;\vec{\nu}^{N=4}_j)$ for different  ground states of filling fractions $\vec{\nu}^{N=4}_j$ with $j=1,2,3,4$. In Fig.~\ref{fig:correlation_energy_W}(a), we show $qW(q;\vec{\nu}^{N=4}_j)/(2n_e\pi)$ versus $q$ at $r_s=1.5$ for a electron gas with $N=4$ components.  The area under each curve yields the total correlation energy for its respective state, as derived from Eq.~\eqref{eq:epsilon_c}. We identify three important features of the $qW(q;\vec{\nu}^{N=4}_j)$ versus $q$ curves: 1) Each curve takes off with a distinct negative slope at $q=0$, the slope being steeper for larger $j$. 2) As $q$ increases, all curves decrease and reach different  minima around $q\approx 2 k_{F}(\vec{\nu}^{N=4}_j)$ where $k_{F}(\vec{\nu}^{N=4}_j)$ are the Fermi wavenumbers of the (equally polarized) states  characterized by filling vector $\vec{\nu}^{N=4}_j$. 
Furthermore, the value at minima becomes increasing negative with increasing $j$. 3) All curves gradually converge  at very large $q$. In what follows, we explain  each of these features.

\begin{figure}[b]
 \includegraphics[width=\columnwidth]{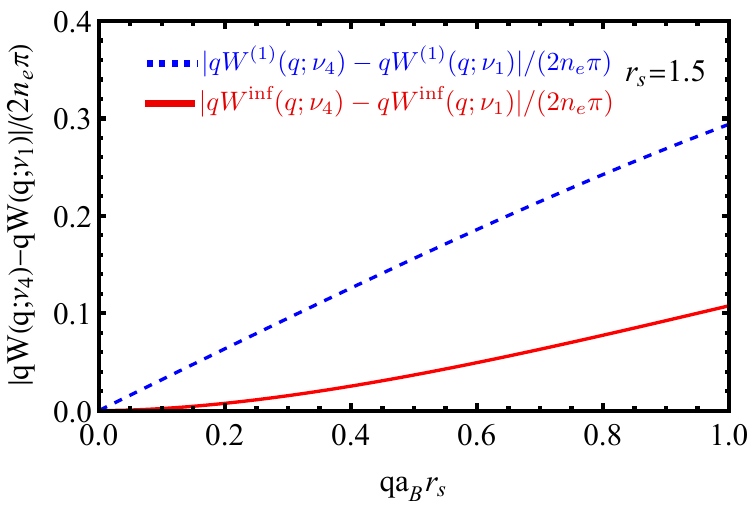}
\center
\caption{  
 Differences of $W^{(1)}(q;\vec{\nu^{N=4}_j})$ and $W^\text{inf}(q;\vec{\nu}^{N=4}_j)$ between the $\vec{\nu}^{N=4}_4$ and $\vec{\nu}^{N=4}_1$ states at small $q$ and $r_s=1.5$ (note $k_F(\vec{\nu}^{N=4}_j)=2/(r_s\sqrt{j})$). The hierarchy for $W(q;\vec{\nu_j})$ at small q is mainly  from the contribution of $W^{(1)}(q;\vec{\nu}^{N=4}_j)$.
 }\label{fig:dW}
\end{figure}

For small values of $q$, it is convenient to separate $W(q;\vec{\nu})$ as given by Eq.~\eqref{eq:W} into two contributions: $W(q;\vec{\nu})=W^{\inf}(q;\vec{\nu})+W^{(1)}(q;\vec{\nu})$. In terms of Feynman diagrams, $W^{\inf}(q;\vec{\nu})=
 \frac{1}{\beta}  \sum_{\omega_n}  \ln \left[1-V_q \Pi_{q,\omega_n}(\vec{\nu})\right]  $ corresponds to the infinite sum of bubble diagrams. However, $W^{(1)}(q,\vec{\nu})= \frac{1}{\beta}  \sum_{\omega_n} V_q\Pi_{q,\omega_n}(\vec{\nu}) $ represents a single bubble diagram (i.e. the exchange energy) and is proportional to $V_q$. For the latter, 
 upon conducting the frequency summation of $\Pi_{q,\omega_n}$, we arrive at the 
 equal-time density correlation function of the non-interacting electron gas. The following analytical result for $W^{(1)}(q;\vec{\nu_j})$ for equally polarized states  $\vec{\nu}^{N=4}_j$ can be thus obtained:
\begin{align}  \label{eq:W2_1}
W^{(1)}(q;\vec{\nu_j})&=\frac{-2}{\pi}\left(\frac{2\pi n}{q}\right) \biggl[\sin^{-1}\left(\frac{qa_Br_s\sqrt{j}}{2}\right)\notag\\&+\frac{qa_Br_s\sqrt{j}}{2}\sqrt{1-\frac{q^2a_B^2r_s^2j}{4}}\biggr].
\end{align}
Let us next take up $W^{\inf}(q;\vec{\nu}^{N=4}_j)$. Performing the analytical continuation  of $W^{\inf}(q;\vec{\nu}^{N=4}_j)$   from  imaginary (Matsubara) frequencies 
to real frequencies  at zero temperature followed by the frequency integration using the residue theorem, we obtain the following formula:
\begin{align} \label{eq:W2_q}
  W^{\inf}(q;\vec{\nu }^{N=4}_j)&=\sum_{r=1}^{P}\hbar \omega_r(q;\vec{\nu}^{N=4}_j) \\
   &\xrightarrow{q\approx 0}
    \sum_{n=1}^{P-1}\hbar \omega_r (q;\vec{\nu})+\hbar\omega_{P}(q),
\end{align}
where  $\{ \omega_r(q;\vec{\nu})\;|\,r=1,2,...P\}$  is the set of  positive roots of the equation 
\begin{equation}\label{eq:1/eps_root}
    1-V_q\Pi_{q,\omega_r}(\vec{\nu})=0.
\end{equation}
Notice that, for small $q$ the largest root corresponds to the plasmon energy, $\hbar\omega_{P}(q)$, making up nearly half of  $W^{\inf}(q;\vec{\nu}^{N=4}_j)$. At small $q$, the plasmon dispersion $ \hbar\omega_{P}(q)=\frac{1}{r_s}\sqrt{2(qa_B)}$ is independent of the state of spin-polarization. For $\vec{\nu}=\vec{\nu}^{N=4}_j$, the contribution of the remaining  roots, i.e. $\sum_{r=1}^{P-1}\hbar \omega_r(q;\vec{\nu}^{N=4}_j)$,  has only very weak dependence on $j$, as shown in Fig.~\ref{fig:dW} \footnote{In the thermodynamic limit, the spacing between the roots $1,2,...P-1$ diminishes, leading to a particle-hole continuum. Consequently, the discrete summation should become an integral.}. This is because the kinetic energy of a particle-hole pair, $\epsilon_{\vec{k}+\vec{q}}-\epsilon_{\vec{k}}$ is dwarfed by their mutual Coulomb interaction $V_q$, at very small $q$. Thus, the roots of Eq.\eqref{eq:1/eps_root} at small $q$ describe a highly collective motion that is insensitive to the spin-polarization of the state.  
Therefore, for small $q$ the dependence on the state $\vec{\nu}^{N=4}_j$ in the RPA correlation energy mainly originates from the first order term  $W^{(1)}(q;\vec{\nu}^{N=4}_j)$. As the number of populated components (i.e. $j$) increases,  $W^{(1)}(q;\vec{\nu}^{N=4}_j)$ decreases, accounting for the behavior of $q W(q,\vec{\nu}^{N=4}_j)$ at small $q$ observed  in Fig.~\ref{fig:correlation_energy_W}.

The wavevectors where the correlation energy substantially favors the paramagnetic state over the symmetry-broken states (namely, the three-quarter metal $j=3$, half-metal $j=2$ and quarter-metal $j=1$) are approximately up to  $2k_{F}(\vec{\nu}^{N=4}_j)=\frac{4}{r_sa_B\sqrt{j}}$. Within this range, the particle-hole susceptibility of the paramagnetic state is significantly enhanced across a broad frequency spectrum up to $\omega\approx 3E_F(\vec{\nu}^{N=4}_4)$.  The most pronounced enhancement of the particle-hole susceptibility is for the $j=4$ paramagnetic case, as shown in Fig.~\ref{fig:chi}. Consequently, when summing over the frequency to compute the correlation energy for a given $q$ (i.e. Eq.~\eqref{eq:W}),  the magnitude of $q W(q;\vec{\nu}^{N=4}_j)/2n_e\pi$ increases in this $q$-range with the number of components $j$.

At very large $q$ such that $V_{q}\Pi_{q,\omega_n} \ll 1$,
the behavior of $W(q;\vec{\nu}^{N=4}_j)$ can be understood from second order perturbation 
theory~\footnote{The infra-red divergence of correlation energy only manifests at the third order in two dimensions because $ E^{(3)}_{\mathrm{corr}} \sim \protect \int^{q_c}_0 d^2  q (V_q)^3 \left[\frac{q^3}{v_F^2 q^2}\right] \sim  \protect \int^{q_c}_0 dq/q$
Here $V_q \sim 1/q $ is the Fourier component of the Coulomb potential, the phase-space $[n_F(\epsilon_k+q)-n_F(\epsilon_k)]^3\sim q^3$ and the particle-hole excitation $\epsilon_{k+q}-\epsilon_k=v_F q$.} In Fig.~\ref{fig:correlation_energy}(d), we show 
$qW^{2nd}(q;\vec{\nu}^{N=4}_j)/(2n_e\pi)$ versus $q$ where $W^{2nd}(q;\vec{\nu_j})$ is given by the following expression:
\begin{align}
W^{2nd}(q,\vec{\nu}^{N=4}_{j}) &= \frac{j^2 V_q^2}{2}\times \int   \frac{d^2k d^2p}{(2\pi)^{4}} \notag\\
&\frac{
n_F(\epsilon_{\vec{k}}) n_F(\epsilon_{\vec{p}})
\bar{n}_F(\epsilon_{\vec{k+q}}) \bar{n}_F(\epsilon_{\vec{p-q}})}{\epsilon_{\vec{k}}+\epsilon_{\vec{p}}-\epsilon_{\vec{k+q}}-\epsilon_{\vec{p-q}}},
\end{align}
where, for the sake of notational simplicity, we have suppressed the dependence of the filling configuration  $\vec{\nu}^{N=4}_j$ of the Fermi-Dirac distribution $n_F(\epsilon)$ and used the notation
$\bar{n}_F(\epsilon)= 1-n_F(\epsilon)$.
Note both $W(q;\vec{\nu}^{N=4}_j)$  and $W^{(2)}(q;\vec{\nu}^{N=4}_j)$ show a very weak  dependence on the number of populated components $j$ at very large $q$. Thus, they do not significantly influence the magnetic phase transition. In addition,  while the second order perturbation theory predicts a single magnetic phase transition, it overestimates the energy difference between the magnetic states and the paramagnetic state. This overestimation primarily arises from the discrepancies at small $q$.

 From the detailed discussion on the partitioning of correlation energy across various wavevectors, it becomes clear that the enhanced particle-hole susceptibility in the paramagnetic state, compared to the symmetry-broken states, plays a pivotal role in favoring the paramagnetic state as the ground state. This enhancement spans a broad spectrum of wavevectors and frequencies.

\section{Comparison with Stoner model}
\label{sec:stoner_model}
\begin{figure}[t]
 \includegraphics[width=\columnwidth]{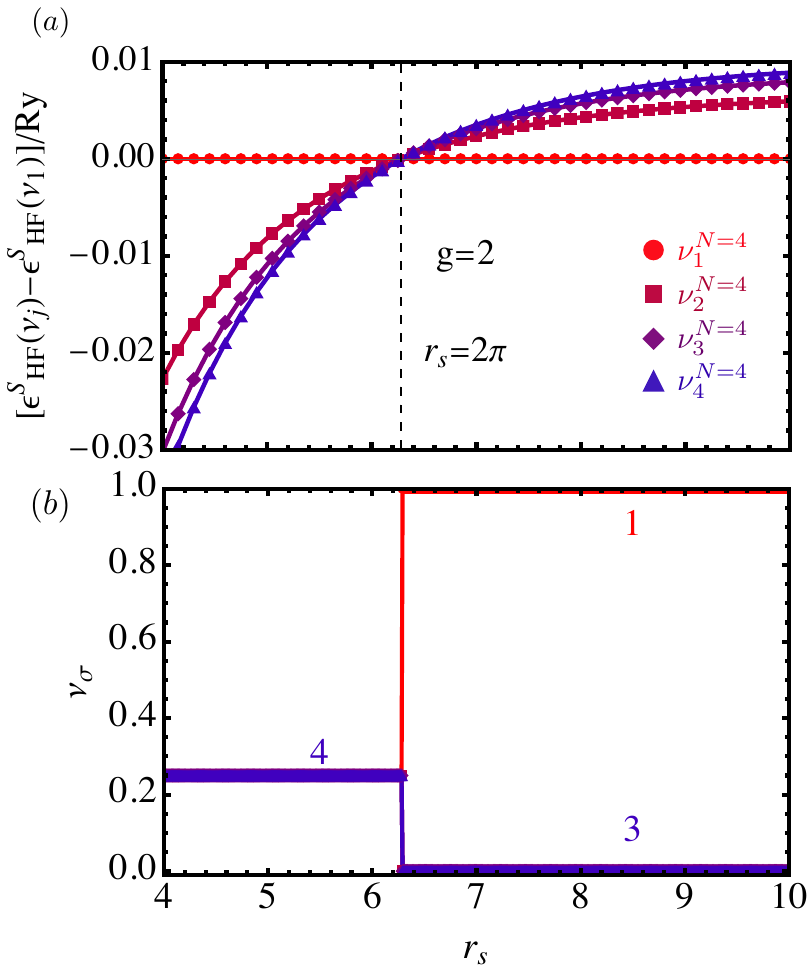}
\center
\caption{
\textbf{(a)} Total energies per particle relative to the fully polarized state calculated for the short range interaction within mean-field theory.
\textbf{(b)} Population of each component derived by minimizing the total energy.
}\label{fig:short}
\end{figure}
 Transitions from the paramagnetic to the ferromagnetic state are often 
 understood using the Stoner model. The latter is based on the assumption that, after accounting for electron correlation, the Coulomb interaction is screened and thus reduced
 to a short-range, density dependent effective interaction~\cite{Ceperly2004,giuliani_vignale_2005} between quasi-particles,
as in Landau's Fermi liquid
theory~\cite{BaymPethick,giuliani_vignale_2005}. The resulting effective Hamiltonian is  treated using mean-field theory. For the two-dimensional Fermi gas considered here,  this approach correctly predicts a  single first-order ferromagnetic transition from the paramagnetic to the fully polarized state as a function of $r_s$. However, as in the HFA described in Sec.~\ref{sec:HF}, it assigns 
the kinetic energy a dominant role  in the energetics of the transition. On the other hand, 
as shown in the previous section, a more accurate treatment of correlation using the RPA shows that,  for the large value of $r_s$ at which the ferromagnetic transition  occurs, the latter is mainly driven  by the competition of the exchange and correlation energies of  different magnetic states, whereas  the kinetic energy which falls off as $r^{-2}_s$ only plays a subdominant role.

The effective   Hamiltonian  of the Stoner model reads:
 \begin{align}
 H_{S} &= H_0 + H_{\mathrm{int}},\\
 H_0 &=\sum_{\vec{p},\sigma} \epsilon_p c^{\dag}_{\vec{p} \sigma} c_{\vec{p}\sigma},\\ 
 \quad
 H_{\mathrm{int}} &= \frac{gr_se^2a_B}{2} \sum_{\vec{p}\vec{k}\vec{q},\sigma < \sigma^{\prime}} c^{\dag}_{\vec{k}+\vec{q},\sigma} c^{\dag}_{\vec{p}-\vec{q},\sigma^{\prime}} c_{\vec{p}\sigma^{\prime}}
 c_{\vec{k}\sigma}.
 \end{align}
 In the mean-field (i.e. Hartree-Fock) approximation, the total energy per particle in atomic units can be written as follows:
 \begin{align}
  \epsilon^S_{HF} = \frac{2}{r_s^2}\sum_\sigma \nu_\sigma^2 + \frac{g}{\pi r_s}\sum_{\sigma < \sigma^{\prime} }\nu_\sigma \nu_{\sigma^\prime}.\label{eq:HFS}
 \end{align}
Squaring  the constraint that fixes the total number of electrons, i.e. $\sum_{\sigma} \nu_{\sigma} = 1$, we have
\begin{equation}
\sum_{\sigma}
\nu^{2}_{\sigma}
=   1 - 2
\sum_{\sigma < \sigma^{\prime}}
\nu_{\sigma} \nu_{\sigma},  
\end{equation}
which allows  to rewrite 
the mean-field   energy in Eq.~\eqref{eq:HFS} as follows:
 \begin{align}
  \epsilon^S_{HF} = \left( \frac{g}{\pi r_s} - \frac{4}{r_s^2} \right) \sum_{\sigma < \sigma^{\prime} }\nu_\sigma \nu_{\sigma^\prime} + \frac{2}{r^2_s}. 
 \end{align}
Thus, for $r_s < 4\pi/g$,
the energy is minimized by 
maximizing the sum $S = \sum_{\sigma< \sigma^{\prime}} \nu_{\sigma}
\nu_{\sigma^{\prime}}$, which happens in the paramagnetic state where $\nu_{\sigma} = 1/N = 1/4$ and $S =  (N-1)/2 = 3/2$. Conversely, for  
$r_s > 4\pi/g$, the energy is
minimized by minimizing $S$, which happens for the fully polarized state where one $\nu_{\sigma}$ is unity and the rest are zero so that $S = 0$. 

The above analysis can be confirmed by direct minimization of the mean-field energy. Fig.~\ref{fig:short}(a) shows the total energy relative to the fully polarized state. The transition point is at $r_s\simeq 2\pi$ (for $g = 2$) and the transition pattern shown in
Fig.~\ref{fig:short}(b) shows a single transition, similar to the RPA to the long-range Coulomb interaction in Sec.~\ref{sec:RPA}. 

However, the energetics of the transition is different from the RPA as seen in Eq.~\eqref{eq:HFS}. Indeed,  for the stoner model, the ferromagnetic transition
is driven by a competition
of the kinetic and mean-field interaction energy, which
are the only two contributions
to the total mean-field energy according to~\eqref{eq:HFS}: As
$r_s$ increases, it becomes advantageous to maximize the mean-field energy even if that requires an increase of the total kinetic energy.  On the other hand, within the RPA, the energetics of the ferromagnetic transition primarily involves a competition between correlation and exchange energies, as the kinetic energy is an order of magnitude smaller at the $r_s\sim 6$ values where the transition occurs.

\section{Summary and Discussions}\label{sec:fin}

In this article, we have studied  density-driven magnetic phase transition of a four component two-dimensional electron gas using both the Hartree-Fock (HF)  and the random-phase approximations (RPA). The total energy versus $r_s$ curve for the three magnetic states (three-quarter metal, half-metal, and fully-polarized metal), as well as the paramagnetic state, varies based on the chosen approximation. In the HF approximation, as $r_s$ increases from small values, the system undergoes a sequence of first-order phase transitions from the paramagnetic state, progressively transitioning to the three-quarter metal,  the half-metal, and ultimately settling in a fully-polarized metal as the ground state. 
However, upon including the correlation energy via RPA, neither the three-quarter metal nor the half-metal are ground states. Instead, the paramagnetic state transition directly into the fully-polarized state.

We have also presented a comparison  between the ferromagnetic transitions of a short-range effective interaction  (Stoner-like) model  and RPA calculations. 
The model is motivated by the screening of the Coulomb interactions that takes place
when electron correlation is 
taken into account. 
Although both models yield  transition patterns that show no cascade and a single transition between fully and equally polarized states, the energetics of  the transition is very different.

Finally, let us apply our simple theoretical model to shed light on the metallic magnetism observed in lightly-doped multilayer graphene. At  low-energies, the physics of  multilayer graphene is described by the electronic degrees of freedom at the two opposite corners of the hexagonal Brillouin zone, (usually termed valleys and labeled with a pseudo-spin index $\tau=\pm1$). Since the  Coulomb interaction is invariant under rotations in both  valley pseudo-spin and spin degrees of freedom, it exhibits an enlarged $SU(4)$ symmetry. However, the band structure  described the single-particle Hamiltonian at different valleys is different.  
Consequently, only a $SU^{\text{spin}}_{{\tau=+}}(2)\times SU^{\text{spin}}_{{\tau=-}}(2)\times U^{\text{valley}}(1)$ symmetry, describing independent spin rotations in the two valleys and valley $\tau^z = N_{+} - N_{-}$ conservation, is realized when considering  together the Coulomb interaction and the band structure. This reduced symmetry means that the Hartree-Fock energy curves ($\epsilon_{HF}$ vs $n_e$) for multilayer graphene electron gases do not exhibit invariance under rotations of the order parameter in the enlarged SU($4$)   spin and valley space. Thus, the energy curves for the three-quarter, half-metal, and quarter-metal states split into two almost degenerate curves. These curves correspond to valley-polarized states (Ising-like) and inter-valley coherent states (XY-like)~\cite{huang2023spin}. The small energy splitting between these valley doublets can be interpreted as a form of magnetic anisotropic energy, which depends on factors such as spin-orbit coupling,  sublattice- and valley-dependent interactions, etc.

Similar to the electron gas with parabolic dispersion electron studied here, magnetic transitions in multilayer graphene electron gases occur within a narrow density window \cite{zhou2021half,zhou2022isospin}. Including the correlation energy, we expect the particle-hole susceptibility of the paramagnetic state to exceed that of the symmetry-broken states. However, due to the complicated multilayer graphene band structure near the Dirac point, the enhancement of the paramagnetic state susceptibility over the magnetic-state susceptibility should have a more complicated form in the $q$-$\omega_n$ plane. Despite these complexities, based on the lessons learned from the study of the $N$-component electron gas described above, the RPA correlation energy between the paramagnetic and symmetry-broken states is expected to overshadow the Hartree-Fock energy differences in the density window where Hartree-Fock predicts magnetic transitions. Thus, we expect the Bohm-Pines RPA correlation energy calculations to lead to a single-phase magnetic transition in multilayer graphene. This prediction, however, contradicts the experimental observations, indicating that the RPA may severely overestimate the magnitude of the correlation energy across the density range where magnetic phase transitions are detected.

In experiments with rhombohedral trilayer graphene \cite{zhou2021half}, the observed symmetry-broken phases are half-metal and quarter-metal, notably without the presence of the three-quarter metal, a pattern also seen in bilayer graphene \cite{zhou2022isospin}. In the HF approximation, we found that the three-quarter metal state emerges as the  ground state with the smallest density interval $1.26<r_s<1.51$, being 
tightly flanked by the more stable paramagnetic and half-metal states. In this 
small density interval, the energy separation to the exited states (i.e. 
paramagnet and  half-metal) is very small. This tight flanking creates a small energy separation to the excited states, suppressing the three-quarter metal even in a simple parabolic electron gas model.

Our research highlights the intricate density dependence of  the total energy among the four competing states. The energy crossings, 
whether between excited states or between excited and ground states, play 
a pivotal role in shaping the magnetic phase diagram. The specific 
approximation chosen to study their mutual interactions, whether RPA
screened interaction, ultra short-range, or long-range Coulomb interaction,  leads to unique patters of energy level crossings. These
findings may serve as a foundational base for further exploration into the 
intricate magnetic phase diagram of multi-component electron gases.

\section{Acknowledgments}

 This work  has been supported by the Agencia Estatal de Investigación (AEI) of the
 Spanish Ministerio de Ciencia e Innovación (MCIN) through Grant No. PID2020-120614GB-I00/AEI/10.13039/501100011033 (ENACT). CLH gratefully acknowledges the hospitaly of the Donostia International Physics Center (DIPC), where this work was started. 
 
\appendix
\section{Correlation in the large $N$ limit}\label{app:LargeN}
\begin{figure}[b]
 \includegraphics[width=\columnwidth]{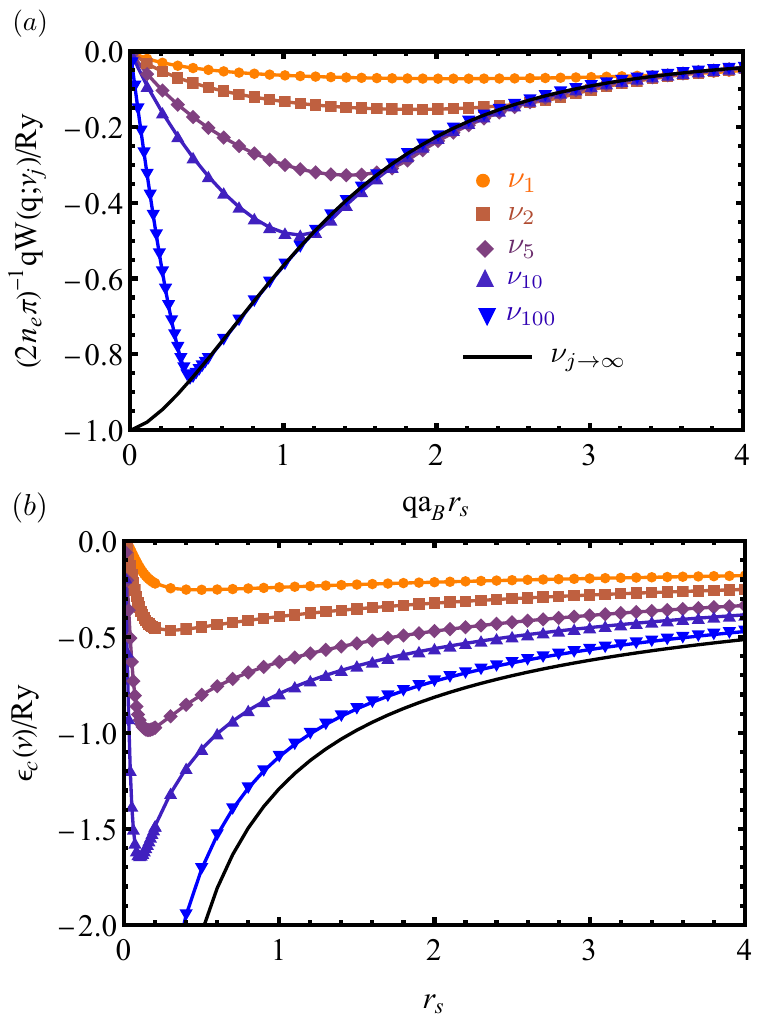}
\center
\caption{(a) Comparison of $q W(q;\vec{\nu}^{N}_j)/(2n_e\pi)$ for $\vec{\nu}^{N=1}_1,
\vec{\nu}^{N=2}_2,
\vec{\nu}^{N=5}_5,
\vec{\nu}^{N=10}_{10}$ and $\vec{\nu}^{N=100}_{100}$ states to the large $N$ limit at $r_s=1.5$ in two dimensions. (b)  Comparison of correlation energies for 
$\vec{\nu}^{N=1}_1,
\vec{\nu}^{N=2}_2,
\vec{\nu}^{N=5}_5,
\vec{\nu}^{N=10}_{10}$, and 
$\vec{\nu}^{N=100}_{100}$ states to the large $N$ limit correlation energy obtained by inserting  Eq.~\eqref{eq:largeNW}.
 into Eq.~\eqref{eq:epsilon_c2}.}\label{fig:lgN}
\end{figure}

 As shown in Fig.~\ref{fig:correlation_energy_W}, the correlation energy function $q W(q;\vec{\nu}^{N}_j)/(2n_e\pi)$ of the different states asymptotically saturates the same function for $q$ large. Such a limit coincides with the large $N$ limit of this function. Indeed, the large $N$ limit of $W(q,\vec{\nu}^{N}_{j\leq N})$ and the resulting correlation energy  can be obtained from the large $N$ limit of Lindhard function for states with $j\leq N$ equally populated components. In this limit,  the effects of the Pauli exclusion principle are weakened, and   the Lindhard function reduces to the density correlation function of a non-interacting Bose gas:
\begin{align}
\Pi_{q,\omega_n}^B=\lim_{j, N\to\infty}j\Pi_{q,\omega_n}(\vec{\nu}^{N}_{j\leq N}) 
=  \frac{-2 n_e \epsilon_{q }}{\omega^2 + \epsilon^2_{q} },
\label{eq:largeNpi}
\end{align}
where $n_e$ is the total electron density and $\omega_n$ is the imaginary (Matsubara) frequency. In this large $N$ limit, the density response of the fermionic system equivalent to that of a Bose-Einstein condensate. This effective ``bosonization''  has been experimentally observed in the density excitations of an ultracold Fermi gas with energent SU($N$) symmetry in the limit of large number of spin components $N$~\cite{PhysRevX.10.041053}.

Here we are concerned with the calculation of the correlation energy at zero temperature for a Fermi gas interacting with the long-range Coulomb interaction. The correlation energy in the RPA (which becomes exact in the limit of large $N$) is given by Eq~\eqref{eq:epsilon_c} and \eqref{eq:W}. Expanding the logarithm using $\ln(1 - x) = -\sum_{n=1}^{+\infty} \frac{x^n}{n}$ we obtain,
\begin{align}
W^B(q) &=  -  \sum_{m=2}^{+\infty}  \frac{V_q^m}{m} \left[ \int_{-\infty}^{\infty} \frac{d\omega}{2\pi} \left[  \Pi_{q,\omega_n}^B \right]^m \right] ,\notag\\
&=   -  \left[  \sum_{m=2}^{+\infty} \frac{\left[-2 n \epsilon_{\vec{q}}  V_q\right]^m}{m} \int_{-\infty}^{\infty}  \frac{d\omega_n}{2\pi}  \frac{1}{\left[ \omega_n^2 + \epsilon^2_{q} \right]^m} \right].
\end{align}
Using the formula for the residue of the higher order poles of arising from $  \Pi_{q,\omega_n} $
\begin{equation}
\int \frac{dx}{\left(x^2 + 1\right)^m} = \frac{\sqrt{\pi}\,\Gamma\left(m-\tfrac{1}{2}\right)}{\Gamma(m)}
\end{equation}
and making an integration variable change to $x = \omega_n/\epsilon_{\vec{q}}$ in the above integral over $\omega_n$, together with $m\Gamma(m) = \Gamma(m+1)$, we arrive at the following series:
\begin{equation}
W^B(q) = -   \left[   \frac{\epsilon_{q}}{2\sqrt{\pi}} \sum_{m=2}^{+\infty}  
 \left[\frac{-2 n   V_q}{\epsilon_{q}}\right]^m  \frac{\Gamma\left(m-\tfrac{1}{2}\right)}{\Gamma(m+1)}\label{eq:finalseries}
\right].
\end{equation}
The series can be resummed by noticing that
\begin{equation}
(1 - z)^{\alpha} = \frac{-1}{2\sqrt{\pi}} \sum_{m=0}^{+\infty} \frac{\Gamma(m-\alpha)}{\Gamma(m+1)} z^m.
\end{equation}
Thus we arrive at the following expression:
\begin{align} \label{eq:largeNW}
W^B(q)  & =   \omega^B_p(q) - \epsilon_{\vec{q}} - n_e V_q 
\end{align}
and
\begin{equation} \label{eq:epsilon_c2}
    \epsilon^B_{c}  = 
\pi (a_Br_s)^2
\int \frac{d^2q}{(2\pi)^2}  W^B(q),\\
\end{equation}
where we have define the frequency:
\begin{equation}
\omega^B_p(q) = \sqrt{\left(2 n_e V_q + \epsilon_{\vec{q}} \right)   \epsilon_{\vec{q}}}
\end{equation}
This frequency corresponds to the zero of the inverse RPA dielectric function in the  large $N$ limit:
\begin{align}
\epsilon^{-1}(q,\omega) =  1 - v(q) \Pi^{B;R}_{q,\omega} 
= \frac{\omega^2 - [\omega^B_p(q) ]^2}{(\omega^+)^2 - \epsilon^2_{\vec{q}}}
\end{align}
Recall that the retarded Lindhard function $\Pi^{B;R}$ is obtained by analytical continuation to real frequency ($\omega^{+} = \omega+i0^{+})$
of the imaginary frequency result in Eq.~\eqref{eq:largeNpi}:
\begin{equation}
\Pi^{B;R}_{q,\omega} = \Pi^{B;R}_{q,\omega_n \to -i \omega^{+}} 
 =  \frac{2n_e \epsilon_{\vec{q}} }{(\omega^+)^2 - \epsilon^2_{\vec{q}}}.
\end{equation}
 Fig.~\ref{fig:lgN} shows the comparison of $q W(q;\vec{\nu}^{N}_{j\leq N})/(2n_e\pi)$ for large $N=j$ and  the correlation energy to the large $N$ limit. $qW(q;\vec{\nu}^{N}_{j\leq N})/(2n_e\pi)$ approaches to the large $N$ limit after the peaks located near $q\sim2k_F(\vec{\nu_j})$.

\section{Generalization to other components}

For $N$ components and fixed $r_s$, there is a family of ground states that are local minima of the HF energy. The filling fractions of such states can be written as a vector $\vec{\nu}_j^N = (\nu_1,\nu_2,\cdots,\nu_j,0,\cdots,0)$, which means they contain $j$ equally-populated  components, $\nu_1,\cdots,\nu_j$, and  $N-j$ totally depopulated components. Note that $\nu_n=1/j$ for all  $n\leq j$. For example, for $N=2$, the two competing states are shown the paramagnetic $\vec{\nu}^{N=2}_{2}$ and fully polarized state $\vec{\nu}^{N=2}_{1}$. For $N=4$, the four competing ground states  in table~\ref{tab:SU(4)}. The following  relation between the total energy 
of ground states states with filling $\vec{\nu}$, denoted $E_0(\vec{\nu})$ below, can be established:
\begin{equation}\label{eq:theorem}
E_0(\vec{\nu}^{N_1}_j) = E_{0}(\vec{\nu}^{N_2}_j).
\end{equation}
for arbitrary $r_s$ and  $N_1, N_2 \ge j$. For example, this means that the total energy of the half-metal phase of the electron gas with $N=4$ components, $\vec{\nu}^{N=4}_{2}$ is identical to the energy of paramagnetic state, $\vec{\nu}_{2}^{N=2}$ in the electron gas with  $N=2$ components. In the HF approximation, Eq.~\eqref{eq:theorem} can be easily established 
for $E_0(\nu) = E^{HF}_0(\vec{\nu}) =  \langle \text{HF}(\vec{\nu})|H|\text{HF}(\vec{\nu})\rangle$. From Eq.~\eqref{eq:pi} and \eqref{eq:phi}, it can be seen that the relation given in Eq.\eqref{eq:theorem} also applies to the correlation energy obtained from the RPA.  Indeed, since the contribution from the depopulated components  to the Lindhard function is zero, the RPA correlation energies for all ground states with $\vec{\nu_{j\leq N}^{N}}$ for any integer $N$ must be equal.

Going beyond the RPA, the exact correlation energy per particle is parameterized by the equal time density correlation function using the method of coupling-constant integration~\cite{giuliani_vignale_2005}:
\begin{equation}
    \epsilon_c = \frac{1}{N_e}\sum_{q\neq0}  \frac{V_q}{2\Omega} \bigg[ \int_0^1   \langle \psi_\lambda |  \rho_{\vec{q}} \rho_{\vec{-q}}| \psi_\lambda \rangle d\lambda -
     \langle \text{HF}|  \rho_{\vec{q}} \rho_{\vec{-q}}| \text{HF} \rangle \bigg],
\end{equation}
where $|\psi_{\lambda}\rangle$ is the ground state of the system with scaled interaction $\lambda V_{q}$ and $\rho_{\vec{q}}=\sum_{\vec{k},\sigma} c^{\dagger}_{\vec{k+q},\sigma}c_{\vec{k},\sigma}$ is the Fourier component of the density operator with wave vector $\vec{q}$. 

Since $|\psi_\lambda\rangle$ can be considered as a perturbed wave function of $|\text{HF}\rangle$ by the scaled Coulomb interaction which does not contain any spin-flip process, the density of the each component in $|\text{HF}\rangle$ and $|\psi_\lambda\rangle$ must be the same. Therefore, when the ground state is void for $\sigma$ particles, we have $c_{k\sigma}|\psi_\lambda\rangle=c_{k\sigma}|\text{HF}\rangle=0$, the correlation energy is only contributed by occupied components and therefore, the exact correlation energy also satisfies Eq.~\eqref{eq:theorem}. This seemingly trivial property allows us to relate the energy versus density curves of an $N$-component system to systems with smaller number of components. 

\bibliographystyle{ieeetr}
\bibliography{reference}

\end{document}